\documentclass{emulateapj}

\usepackage{timesfonts}

\usepackage{amsmath}

\def\M324{\ensuremath{M_{g,324}}}
\def\160d{160~deg$^2$}
\def\ergspercm{erg~s$^{-1}$~cm$^{-2}$}
\def\ergs{erg~s$^{-1}$}

\def\***#1{{\scshape***#1***}}

\makeatletter
\renewcommand\normalsize{%
   \@setfontsize\normalsize\@xpt\@xiipt
   \abovedisplayskip=0.25\baselineskip
   \belowdisplayskip=\abovedisplayskip
   \abovedisplayshortskip=\abovedisplayskip
   \belowdisplayshortskip=\belowdisplayskip
   \let\@listi\@listI}
\makeatother

\begin{document}

\parskip=0pt plus 0.1pt

\lefthead{COSMOLOGICAL PARAMETERS FROM CLUSTER EVOLUTION}
\righthead{VIKHLININ ET AL.}

\journalinfo{ApJ in press; astro-ph/0212075}
\submitted{Submitted to ApJ 11/29/02; accepted 2/18/03; astro-ph/0212075}

\title{Cosmological constraints from evolution of \\ cluster baryon mass
  function at $z\sim0.5$}

\author{A. Vikhlinin\altaffilmark{1,2}, A.\
Voevodkin\altaffilmark{2}, C.~R.~Mullis\altaffilmark{3},
L.~VanSpeybroeck\altaffilmark{1}, H.~Quintana\altaffilmark{4},
B.~R.~McNamara\altaffilmark{5}, I.~Gioia\altaffilmark{6},
A.~Hornstrup\altaffilmark{7}, J.~P.~Henry\altaffilmark{8},
W.~R.~Forman\altaffilmark{1}, C.~Jones\altaffilmark{1}}

\altaffiltext{1}{Harvard-Smithsonian Center for Astrophysics, 60 Garden St.,
  Cambridge, MA 02138; avikhlinin@cfa.harvard.edu}
\altaffiltext{2}{Space Research Institute, Moscow, Russia}
\altaffiltext{3}{European Southern Observatory, Karl-Schwarzschild-Stra\ss e
  2, D-85748 Garching, Germany}
\altaffiltext{4}{Dpto.\ de Astronomia y Astrofisica, Pontificia Universidad
  Catolica, Casilla 306, 22 Santiago, Chile}
\altaffiltext{5}{Department of Physics and Astronomy, Ohio University,
  Athens, OH 45701}
\altaffiltext{6}{Istituto di Radioastronomia CNR, Via P. Gobetti, 101,
  I-40129, Bologna, Italy}
\altaffiltext{7}{Danish Space Research Institute, Juliane Maries Vej 30,
2100  Copenhagen O, Denmark}
\altaffiltext{8}{University of Hawaii, Institute for Astronomy, 2680
  Woodlawn Drive, Honolulu, HI 96822}

\begin{abstract}
  
We present a new method for deriving cosmological constraints based on the
evolution of the baryon mass function of galaxy clusters, and implement it
using 17 distant clusters from our \160d\ \emph{ROSAT} survey.  The method
uses the cluster baryon mass as a proxy for the total mass, thereby avoiding
the large uncertainties of the $M_{\text{tot}}-T$ or $M_{\text{tot}}-L_X$
relations used in all previous studies. Instead, we rely on a well-founded
assumption that the $M_b/M_{\text{tot}}$ ratio is a universal quantity,
which should result in a much smaller systematic uncertainty. Taking
advantage of direct and accurate \emph{Chandra} measurements of the gas
masses for distant clusters, we find strong evolution of the baryon mass
function between $z>0.4$ and the present. The observed evolution defines a
narrow band in the $\Omega_m-\Lambda$ plane,
$\Omega_m+0.23\Lambda=0.41\pm0.10$ at 68\% confidence, which intersects with
constraints from the Cosmic Microwave Background and supernovae~Ia near
$\Omega_m=0.3$ and $\Lambda=0.7$.

\end{abstract}


\section{Introduction}

The growth of large scale structure in the near cosmological past is very
sensitive to the value of the density parameter, $\Omega_m$, and, weakly, to
the cosmological constant, $\Lambda$ (see, among others, Sunyaev 1971,
Peebles 1980, Carroll, Press \& Turner 1992). One of the strongest
manifestations of the growth of density perturbations is the formation of
virialized objects such as clusters of galaxies (Press \& Schechter
1974). Therefore, the evolution of the cluster mass function provides a
sensitive cosmological test (e.g.\ Evrard 1989 and later works).

Present-day theory provides machinery for accurate predictions for the
cluster mass function at any redshift, in any cosmology (see a brief review
in \S\,\ref{sec:theory}). Most of the difficulties with using cluster
evolution as a cosmological probe are on the observational side.  The
foremost problem is the necessity to relate cluster masses with other
observables, such as the X-ray temperature. Even for the few brightest,
best-studied clusters, the total mass within the virial radius --- the
quantity generally used in theory --- is measured presently with large
uncertainties (see, e.g., Markevitch \& Vikhlinin 1997 and Fischer \& Tyson
1997 for representative uncertainty estimates for the X-ray hydrostatic and
weak lensing methods, respectively). To measure virial masses in a large,
complete sample of clusters is presently impractical. Therefore, one usually
resorts to other, more easily measurable quantities, which exhibit scaling
relations with the total mass. The most widely used observable is the
temperature of the intracluster medium. It is expected that $T$ is related
to the virial mass as $M\propto T^{3/2}$, and the available observations
generally support this law (Nevalainen, Markevitch \& Forman 2000; Horner,
Mushotzky \& Scharf 1999; Finoguenov, Reiprich \& B\"ohringer 2001). To
normalize this relation, virial masses still have to be measured in a
representative sample of clusters. Any systematic uncertainties in the
cluster mass measurements translate into the same uncertainty in the
normalization of the $M-T$ relation.

The X-ray temperature functions (XTF) for both nearby and distant clusters
have been used as a proxy for the mass function by a number of authors,
using the normalization of the $M-T$ relation from different sources. An
analysis of a statistically complete local XTF was first performed by Henry
\& Arnaud (1991). The analysis of the local XTF yields the normalization
($\sigma_8$), in combination with $\Omega_m$, and slope of the power spectrum
of density fluctuations on cluster scales. This type of study has since been
applied to ever improving observational data (see Evrard et al.\ 2002 for a
recent summary).

A first cosmological measurement using the evolution of the XTF at redshifts
greater than zero is due to Henry (1997), who derived $\Omega_m \approx
0.5\pm0.15$. A similar analysis was performed by Eke et al.\ (1998), Donahue
\& Voit (1999), Henry (2000), Blanchard et al.\ (2000), with updates on both
observational and theoretical sides as well as with different normalizations
of the $M-T$ relation.  We also note the studies by Reichart et al.\ (1999)
and Borgani et al.\ (2001) who modeled the X-ray luminosity functions for
distant clusters using an empirical $L-T$ relation and the usual theory for
the XTF evolution.

Theoretical prediction for the number density of clusters with a given
temperature is exponentially sensitive to the normalization of the $M-T$
relation.  A 30\% change in the normalization of the $M-T$ relation results
in a 20\% change in the determination of the $\sigma_8$ parameter and shifts
an estimate of $\Omega_m$ by a factor of 1.3--1.4 (e.g., Rosati, Borgani \&
Norman 2002). Given that $\pm50\%$ is a fair estimate of the present
systematic uncertainties in the $M-T$ normalization, the impact of this
relation on cosmological parameter determination can be very significant.

In modeling the high-redshift temperature function, the theoretically
expected evolution of the $M-T$ relation is usually assumed: $M\propto
T^{3/2}/(1+z)^{3/2}$. It would be very important to verify this evolution at
high $z$. Unfortunately, this seems to push the limits of current
observational techniques. The evolution of other scaling relations can be
also quite important. For example, the $L-T$ relation is required for
computing the volume of X-ray flux limited surveys, and for estimating
cluster masses, when the X-ray luminosity function is used as a proxy for
the mass function. Most of the previous studies have assumed that the $L-T$
relation does not evolve, however recent \emph{Chandra} measurements
indicate otherwise (Vikhlinin et al.\ 2002).

Finally, most of the available cosmological studies are based on the distant
cluster sample from the \emph{Einstein} Extended Medium Sensitivity Survey
(Gioia et al.\ 1990, Henry et al.\ 1992). The completeness of this survey at
high redshifts is often questioned (Nichol et al.\ 1997; Ebeling et al.\
2000; Lewis et al.\ 2002). Although we believe that the EMSS is essentially
complete (as confirmed by good agreement between the XLFs from the EMSS and
several \emph{ROSAT} surveys --- Gioia et al.\ 2001, Vikhlinin et al.\ 2000,
Mullis et al., in preparation), it is arguably important to try another
completely independent sample for the cosmological studies based on cluster
evolution.

The purpose of this paper is to apply the evolutionary test to a new sample
of distant, $z>0.4$, clusters derived from the 160~deg$^{2}$ \emph{ROSAT}
survey (Vikhlinin et al.\ 1998), relying on the evolution of the cluster
scaling relations as measured by \emph{Chandra}, and using a modeling
technique which does not rely upon the total mass measurements to normalize
the $M-T$ or $M-L$ relations and therefore bypasses many of the
uncertainties mentioned above.

The proposed method relies on the assumption that the baryon fraction within
the virial radius in clusters should be close to the average value in the
Universe, $f_{b,U} = \Omega_b/\Omega_m$. This is expected on general
theoretical grounds (White et al.\ 1993) because gravity is the dominant
force on cluster scales. The universality of the baryon fraction in massive
clusters is supported by cosmological numerical simulations (e.g.\ Bialek,
Evrard \& Mohr 2001) and by most available observations (most recently,
Allen et al.\ 2002). In principle, the equality $f_b = \Omega_b/\Omega_m$
can be used to determine cosmological parameters either from the absolute
measurements of $f_b$ in the nearby clusters (White et al.\ 1993) or from
the apparent redshift dependence of $f_b$ measurements in the high-redshift
clusters (Sasaki 1996, Pen 1997). Both methods, however, require an accurate
observational determination of the total mass in individual clusters, which
we would like to avoid.

Unlike the total mass, the baryon mass in clusters is relatively easily
measured to the virial radius (Vikhlinin et al.\ 1999). To first order, the
baryon and total mass are trivially related, $M_b = M\,\Omega_b/\Omega_m.$ If
this is the case, the relation between the cumulative total mass function,
$F(M)$, and the baryon mass function, $F_b(M_b)$, is also trivial,
\begin{equation}\label{eq:f(m_b):f(m):ideas}
  F_b(M_b) = F\left(\Omega_m/\Omega_b\, M_b\right).
\end{equation}
The average baryon density in the Universe is accurately given by
observations of the light element abundances and Big Bang Nucleosynthesis
theory (Burles, Nollet \& Turner 2001). Therefore, we have everything needed
to convert the theoretical model for the total mass function, $F(M)$, to the
prediction for a directly measurable baryon mass function --- to compute
$F(M)$ one must choose a value of $\Omega_m$, which fixes the scaling
between the total and baryon mass. The conversion of the total to the baryon
mass function --- at least to a first approximation --- can be expressed
through the model parameters, and therefore $F_b(M_b)$ can be modeled with
the usual set of parameters, $\Omega_m$, $\sigma_8$, $n$, $h$, and at a high
redshift --- $\Lambda$. We will show below that the evolution of the baryon
mass function provides robust constraints on a combination of $\Omega_m$ and
$\Lambda$. Modeling of the baryon mass function at $z=0$ provides a
measurement of the power spectrum normalization, $\sigma_8$, and of the
shape parameter $\Gamma=\Omega_m\,h$, with degeneracies that differ from
those given by the temperature function (Voevodkin \& Vikhlinin, in
preparation).

Theoretical models of the cluster mass function usually consider masses
measured within radii defined by certain values of the density contrast,
$\delta = 3 M(r)/(4\pi\,r^3)/\rho$, where $\rho$ is either the critical
density or mean density of the Universe. The assumption that the baryon
fraction in clusters is universal at sufficiently large radii allows one to
determine the overdensity radii $r_\delta$ without measuring the total
mass. Indeed, the baryon and matter overdensities are then equal,
$\delta_b=\delta$, and the baryon overdensity is defined by a baryon mass
$M_b(r)$ and the mean baryon density in the Universe, known from the
BBN. One of the most accurate theoretical models for the mass function
(Jenkins et al.\ 2001) uses masses corresponding to the overdensity
$\delta=324$ relative to the mean density of the Universe at the redshift of
observation. The matching baryon masses, $M_{b,324}$ are easily measured
using \emph{ROSAT} data for nearby clusters (Voevodkin, Vikhlinin \&
Pavlinsky 2002a, VVPa hereafter), and \emph{Chandra} data for distant
clusters (Vikhlinin et al.\ 2002).

The paper is organized as follows. In \S\,\ref{sec:data}, we discuss the
distant cluster sample and the gas mass measurements in the distant and
nearby clusters. The baryon mass function for distant clusters is derived in
\S\,\ref{sec:mfun:meas}. The theory for modeling these data is reviewed in
\S\,\ref{sec:theory}. We describe our fitting procedure in
\S\,\ref{sec:fitting}, and present the derived constraints on $\Omega_m$ and
$\Lambda$ in \S\,\ref{sec:omega:lambda:results}.

All cluster parameters are quoted for $h=0.65$, $\Omega_m=0.3$, and
$\Lambda=0.7$. The X-ray fluxes and luminosities are in the 0.5--2 keV
energy band. Measurement uncertainties are $1\sigma$.

\section{Data}
\label{sec:data}

\begin{table*}
\def\arraystretch{1.2}
\def\j{\phantom{1}}
\caption{X-ray data and baryon masses for the distant \160d\ clusters}\label{tab:mfun:sample}
{\centering
\footnotesize
\centerline{
\begin{tabular}{p{3.2cm}ccccccc}
\hline
\hline
\raisebox{-8pt}[0pt][0pt]{Cluster} & \raisebox{-8pt}[0pt][0pt]{$z$} &
$f_{0.5-2}$ (\emph{ROSAT})  & $L_{0.5-2}$ & $M_{g,324}$  & $M_{b,324}$ & $f_{0.5-2}$ (\emph{Chandra}) \\[1pt]
             &   & (\ergspercm) &    \multicolumn{1}{c}{(\ergs)} &
             $(10^{13}\, M_\odot)$ &  $(10^{13}\, M_\odot)$ &
            (\ergs) 
\\
\hline
CL\,1416+4446\dotfill &0.400&$4.04\times10^{-13}$&$2.5\times10^{44}$&$ 8.0\pm2.3$  &$ 9.8\pm2.8$  &$4.51\times10^{-13}$\\
CL\,1701+6414\dotfill &0.453&$3.86\times10^{-13}$&$2.9\times10^{44}$&$10.4\pm2.8$\j&$12.5\pm3.4$\j&$4.14\times10^{-13}$\\
CL\,1524+0957\dotfill &0.516&$3.04\times10^{-13}$&$2.7\times10^{44}$&$ 9.3\pm2.3$  &$11.2\pm2.8$\j&$2.81\times10^{-13}$\\
CL\,1641+4001\dotfill &0.464&$2.94\times10^{-13}$&$2.3\times10^{44}$&\nodata       &$ 8.3\pm2.9$  &\nodata\\
CL\,0030+2618\dotfill &0.500&$2.43\times10^{-13}$&$2.3\times10^{44}$&$ 6.6\pm2.7$  &$ 8.1\pm3.3$  &$2.51\times10^{-13}$\\
CL\,1120+2326\dotfill &0.562&$2.13\times10^{-13}$&$2.2\times10^{44}$&$ 6.6\pm1.5$  &$ 8.2\pm1.9$  &$1.88\times10^{-13}$\\
CL\,1221+4918\dotfill &0.700&$2.06\times10^{-13}$&$4.1\times10^{44}$&$11.3\pm2.1$\j&$13.6\pm2.6$\j&$2.25\times10^{-13}$\\
\hline                                                                                            
CL\,0853+5759\dotfill &0.475&$1.98\times10^{-13}$&$1.6\times10^{44}$&\nodata       &$ 6.3\pm2.2$  &\nodata\\
CL\,0522--3625\dotfill&0.472&$1.84\times10^{-13}$&$1.5\times10^{44}$&\nodata       &$ 5.9\pm2.1$  &\nodata\\
CL\,1500+2244\dotfill &0.450&$1.78\times10^{-13}$&$1.3\times10^{44}$&\nodata       &$ 5.5\pm1.9$  &\nodata\\
CL\,0521--2530\dotfill&0.581&$1.76\times10^{-13}$&$2.3\times10^{44}$&\nodata       &$ 7.2\pm2.5$  &\nodata\\
CL\,0926+1242\dotfill &0.489&$1.67\times10^{-13}$&$1.5\times10^{44}$&\nodata       &$ 5.7\pm2.0$  &\nodata\\
CL\,0956+4107\dotfill &0.587&$1.56\times10^{-13}$&$2.0\times10^{44}$&\nodata       &$ 6.6\pm2.3$  &\nodata\\
CL\,1216+2633\dotfill &0.428&$1.54\times10^{-13}$&$1.0\times10^{44}$&\nodata       &$ 4.6\pm1.6$  &\nodata\\
CL\,1354--0221\dotfill&0.546&$1.45\times10^{-13}$&$1.6\times10^{44}$&\nodata       &$ 5.8\pm2.0$  &\nodata\\
CL\,1117+1744\dotfill &0.548&$1.44\times10^{-13}$&$1.6\times10^{44}$&\nodata       &$ 5.8\pm2.0$  &\nodata\\
CL\,1213+0253\dotfill &0.409&$1.43\times10^{-13}$&$0.9\times10^{44}$&\nodata       &$ 4.2\pm1.5$  &\nodata\\
\hline
\end{tabular}
}
\par

%                    z     mb  emb     L      mg  emg
% cl1416p4446      0.400  9.78 2.82   2.5    8.03 2.32
% cl1701p6414      0.453 12.50 3.37   2.9   10.39 2.80
% cl1524p0957      0.516 11.22 2.75   2.7    9.27 2.27
% cl1641p4001      0.464  8.26 2.89   2.3     ...  ...   
% cl0030p2618      0.500  8.13 3.34   2.3    6.61 2.71
% cl1120p2326      0.562  8.17 1.90   2.2    6.64 1.54
% cl1221p4918      0.700 13.56 2.55   4.1   11.30 2.13

% cl0853p5759      0.475  6.30 2.21   1.6
% cl0522m3625      0.472  5.92 2.07   1.5
% cl1500p2244      0.450  5.47 1.91   1.3
% cl0521m2530      0.581  7.18 2.51   2.3
% cl0926p1242      0.489  5.74 2.01   1.5
% cl0956p4107      0.587  6.60 2.31   2.0
% cl1216p2633      0.428  4.64 1.63   1.0
% cl1354m0221      0.546  5.81 2.04   1.6
% cl1117p1744      0.547  5.81 2.03   1.6
% cl1213p0253      0.409  4.16 1.46   0.9

\medskip

\begin{minipage}{0.99\linewidth}
\footnotesize

All distance-dependent quantities are derived assuming $\Omega_m=0.3$,
$\Lambda=0.7$, and $h=0.65$. For different values of the Hubble constant,
masses scale as $h^{-2.25}$. The luminosities are determined using the
\emph{ROSAT} fluxes, not the \emph{Chandra} measurements. The baryon mass
(i.\,e.\ gas+stars) is estimated from the gas mass
using~(\ref{eq:mgas:mbar}).  For CL\,1641+4001 and all clusters with
$f<2\times10^{-13}$~\ergspercm, the values of $M_{b,324}$ are estimated from
the $L-M_b$ relation.

\end{minipage}
\par
}
\end{table*}

\subsection{Distant Cluster Sample}

The sample of distant clusters used in this work is derived from our \160d\
survey (Vikhlinin et al.\ 1998). This survey is based on the X-ray detection
of serendipitous clusters in the central parts of a large number of the
\emph{ROSAT} PSPC observations. The spatial extent of the X-ray sources is
used as a primary selection criterion, but it is verified by independent
surveys (Perlman et al. 2002) that above the advertised flux limits, we do
not miss any clusters because of the limited angular resolution of
\emph{ROSAT}. Therefore, for all practical purposes the \160d\ cluster
sample can be considered as flux-limited.

The optical identification of the entire X-ray sample is now complete and
redshifts of all clusters are spectroscopically measured (Mullis et al., in
preparation). Optical identifications show that our X-ray selection is very
efficient. In the entire sample, only 20 of the 223 detected X-ray cluster
candidates lack a cluster identification. Above the median flux of the
survey, $f>1.4\times10^{-13}$~\ergspercm, 111 of 114 sources (or 98\%) were
confirmed as clusters. Therefore, the \160d\ survey is in effect purely
X-ray selected.

Our X-ray detection procedure is fully automatic, and therefore all
essential statistical characteristics of the survey can be derived from
Monte-Carlo simulations. We have performed very extensive simulations to
calibrate the process of detection of clusters with different fluxes and
radii. This gave an accurate determination of the survey sky coverage as a
function of flux (Vikhlinin et al.\ 1998), and therefore of the survey
volume for any combination of cosmological parameters.

\subsection{Chandra Observations}

As of Fall 2002, \emph{Chandra} has observed 6 clusters from the \160d\
sample. These objects almost complete a flux-limited subsample,
$f>2\times10^{-13}$~\ergspercm, of objects at $z>0.4$ (7 total), and
therefore they are suitable for statistical studies. \emph{Chandra}
observations of these clusters are discussed in Vikhlinin et al.\
(2002). Briefly, the quality of the X-ray data allows a measurement of the
cluster temperature to a 10\% accuracy, and of the gas mass at the virial
radii with a $20\%$ uncertainty. For the purposes of this paper, we need
only the measurements of the gas mass.

The \emph{Chandra} observations of the distant \160d\ clusters as well as
those from the EMSS and RDCS (Rosati et al.\ 1998) samples show that scaling
relations between the X-ray luminosity, temperature, and the gas mass
significantly evolve with redshift (Vikhlinin et al.\ 2002). In particular,
the $L-M_g$ relation evolves so that the gas mass for a fixed luminosity is
proportional to $(1+z)^{-1.8\pm0.25}$ but the slope and scatter in the
high-redshift relation are the same as in the local relation (Voevodkin,
Vikhlinin \& Pavlinsky 2002b, VVPb hereafter). This allows an estimate of
the gas mass, with a $\pm35\%$ uncertainty, from the easily measured X-ray
flux.

The $L-M_g$ relation allows us to extend the estimate of the gas mass to all
high-redshift clusters that do not have \emph{Chandra}
observations. However, we will use only clusters with
$f>1.4\times10^{-13}$~\ergspercm. At $z=0.4$, this flux limit corresponds
roughly to $T=3$~keV. It is possible that for much less massive clusters,
our basic assumption of the universality of the baryon fraction may not hold
due to, e.\,g., preheating of the intracluster medium (e.g., Bialek et al.\
2001). The median redshift for the selected distant clusters is $z=0.55$.

The measurements and estimates of the baryon mass in distant clusters are
listed in Table~\ref{tab:mfun:sample}. The measured masses were derived
using direct deprojection of the X-ray surface brightness profiles. The mass
uncertainties were obtained by error propagation in the process of
deprojection and so correctly reflect the statistical noise in the surface
brightness profiles near $r_{324}$. The only object with \emph{Chandra} data
which was not used by Vikhlinin et al.\ (2002), is CL\,0030+2618. This
cluster was observed very soon after the \emph{Chandra} launch while the
detector parameters did not yet reach their nominal values. These data are
unusable for spectral (temperature) measurements, but the imaging analysis
which yields the gas mass can be performed without any complications.

The baryon masses were estimated from the $M_b-L$ correlation for
approximately 60\% of clusters in our distant sample. Both quantities in the
$M_b-L$ correlation are easily and directly measured, and so there is no
systematic uncertainty in the derivation of normalization, or slope, or
evolution of this relation.

Finally, we remark on the excellent agreement of the X-ray fluxes for the
\160d\ survey clusters derived from \emph{ROSAT} and \emph{Chandra}
data. For the 6 clusters in common, the \emph{ROSAT} fluxes deviate by no
more than 13\%, always within the statistical uncertainties. The average
ratio of the \emph{ROSAT} and \emph{Chandra} fluxes is $0.99\pm0.06$.

\subsection{Summary of the Low-Redshift Results}

The baryon mass measurements in a large sample of low-redshift clusters, as
well as the determination of the baryon mass function at $z=0.05$ was
reported in VVPa,b. These results provide a low-redshift reference point for
our measurement of cluster evolution, so a brief summary is in order.

VVP have used a flux-limited subsample of the low-redshift clusters detected
in the \emph{ROSAT} All-Sky Survey. The initial object selection was
performed using the HIFLUGCS sample (Reiprich \& B\"ohringer 2001). The
X-ray fluxes were re-measured and 52 clusters were selected in the redshift
interval $0.01<z<0.25$ and with $f_x>1.4\times10^{-11}$~\ergspercm. The
median redshift of this sample is $z=0.05$. Most of these clusters have
\emph{ROSAT} PSPC pointed observations which were used to measure the gas
mass (the procedure is described in VVPa and Vikhlinin, Forman \& Jones
1999).

Using a subsample of clusters with published optical measurements, a tight
correlation between the gas mass and optical luminosity of the cluster has
been established. This relation allows an estimate of the baryon (i.\,e.\
gas+galaxies but excluding intergalactic stars) mass of all clusters from
the X-ray data alone:
\begin{equation}\label{eq:mgas:mbar}
  M_{b,324} = M_{g,324}
  \times\left[1.100+0.045\left(\frac{M_{g,324}}{10^{15}\,M_\sun}\right)^{-0.5}\right], 
\end{equation}
where $M_{b,324}$ is the baryon mass corresponding to the overdensity
$\delta=324$, and $M_{g,324}$ is the gas mass corresponding to $\delta=324$
without accounting for the stellar mass. The stellar contribution to the
baryon mass is small, but non-negligible, of order 10--15\% for massive
clusters (this is consistent with the estimates in Fukugita, Hogan \&
Peebles 1998). This is estimated with approximately 25\% uncertainty
(i.\,e.\ $\pm3\%$ of the total baryon mass). Assuming that the ratio of
stellar and gas mass does not evolve at high $z$, i.\,e.\ that the galaxies
neither confine the intracluster gas nor lose their mass significantly due
to stellar winds, equation~(\ref{eq:mgas:mbar}) can be used to estimate the
baryon mass in distant clusters (our results are not sensitive to this
assumption).

VVPb have derived the low-redshift baryon mass function (reproduced
below). The measurement uncertainties in this mass function were determined
using a technique which accounts for the Poisson noise as well as the
individual $M_b$ measurement uncertainties (see also Appendix).

\begin{figure*}
\centerline{\includegraphics[width=0.49\linewidth]{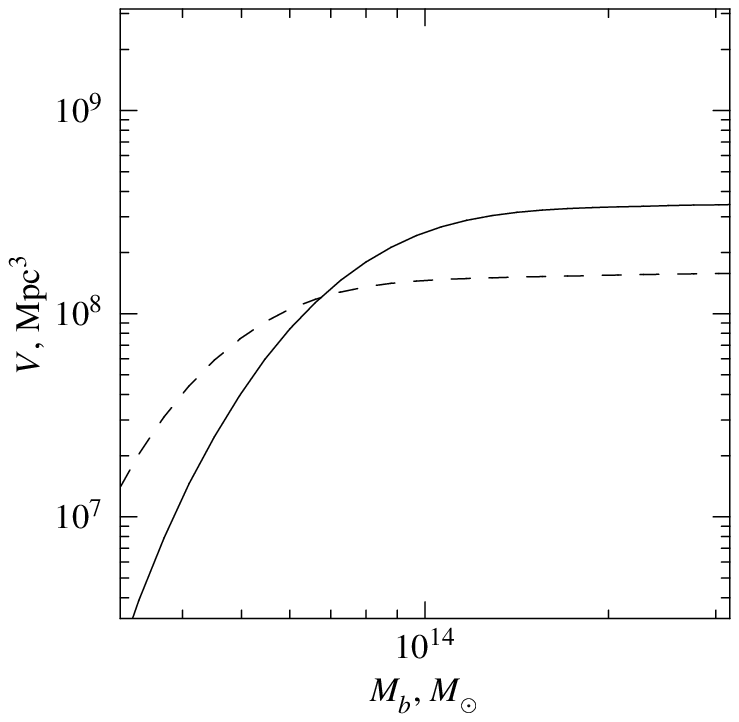}\hfill
\includegraphics[width=0.49\linewidth]{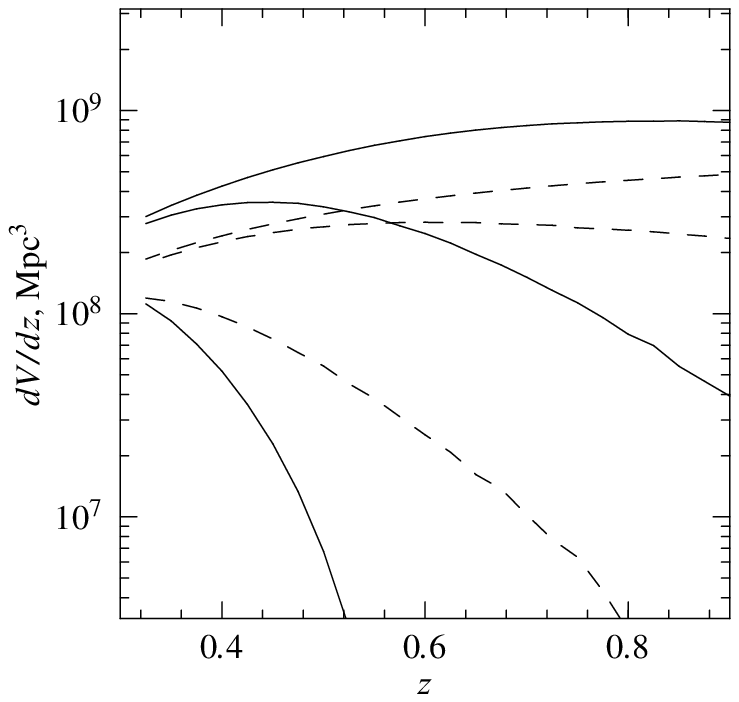}}
 \vskip -1.0\baselineskip
\caption{\emph{Left:} Comoving volume of the \160d\ survey for $0.4<z<0.8$ as a function of the cluster baryon
  mass. \emph{Right}: Volume per unit redshift interval for three values of
  $M_b$ --- 0.25, 0.5, and $1\times10^{14}\,M_\sun$ (from bottom to top). On
  each panel, the solid lines are for the $\Omega_m=0.3$, $\Lambda=0.7$
  cosmology, and dashed lines are for $\Omega_m=1$, $\Lambda=0$. The volumes
  are computed for a limiting X-ray flux of
  $f_{\text{min}}=1.4\times10^{-13}$~\ergspercm.}
\label{fig:mfun:volume}
\end{figure*}

To summarize, the baryon mass function for the low-redshift clusters is
measured reliably. By itself, it provides constraints on the shape and
normalization of the power spectrum of the density perturbations, i.e.\ on the parameters $\sigma_8$ and $\Gamma=\Omega_m\,h$. These
results will be reported in Voevodkin \& Vikhlinin (in preparation). In the
present Paper, we restrict ourselves to the cosmological constraints derived
from the \emph{evolution} of the mass function between $z>0.4$ and
$z\simeq0$.

\section{Baryon mass function at $z>0.4$}
\label{sec:mfun:meas}

In this section, we use the \emph{Chandra} observations to derive the baryon
mass function for distant clusters. First, we describe the computations of
the survey volume as a function of mass. 

\subsection{Survey Volume}
\label{sec:mfun:volume}

We consider the mass function for $0.4\leq z\leq0.8$. Since our cluster
sample is selected by X-ray flux, the computation of the surveyed volume as
a function of mass is only possible if there is a correlation between the
baryon mass and X-ray luminosity. Such a correlation is observed (VVPb,
Vikhlinin et al.\ 2002),
\begin{equation}\label{eq:volume:m:ml}
  M_b \propto (1+z)^{A_{ML}} L_X^{0.83}
\end{equation}
with a 42\% scatter in luminosity and 35\% scatter in mass; $L_X$ is the
total luminosity in the 0.5--2~keV band.

In the case of zero scatter in the $M_b-L$ relation, the volume is
computed as
\begin{equation}\label{eq:volume:m1}
  V = \int_{z_{\text{min}}}^{z_{\text{max}}}
  A(f)\;\frac{dV}{dz} \;dz, \;\; f=\frac{C\,M_b^{1.20} (1+z)^{-1.20A_{ML}}\,S(z)}{4\pi\,
      d_L^2(z)}
\end{equation}
where $A(f)$ is the survey area as a function of flux, $dV/dz$ is the
cosmological dependence of the comoving volume, equation
(\ref{eq:volume:m:ml}) is rewritten as $L=C\, M^{1.20} (1+z)^{-1.20A_{ML}}$,
and $S(z)$ is a correction for redshifting of the bandpass (equivalent to
the K-correction in the optical astronomy). Note that for the non-evolving
$L-T$ relation, $S(z)$ is independent of cosmology. The corrections due to
the observed rate of evolution in this relation (Vikhlinin et al.\ 2002) are
small, $\sim 5\%$ in volume, and they were ignored.

Introducing the scatter in the $M_b-L$ relation with the log-normal
distribution (see VVPb), equation (\ref{eq:volume:m1}) becomes
\begin{equation}\label{eq:volume:m}
  \begin{split}
    V = &\frac{1}{(2\pi)^{1/2}\sigma} \int_{z_{\text{min}}}^{z_{\text{max}}}
    \int_{-\infty}^\infty \exp\left(-\lg^2
      y/2\sigma^2\right) \,d\lg y\; \times \\
    & \;\;\;\times 
    A\left(y\;\frac{C\,M^{1.20} (1+z)^{-1.20A_{ML}}\,S(z)}{4\pi\,
        d_L^2(z)}\right)\;\frac{dV}{dz} \;dz,
    \end{split}
\end{equation}
where $\sigma=0.173$ is the observed \emph{rms} log-scatter in the local
mean $M_b-L$ relation (VVPb). Numerical integration of this equation gives
the survey volume as a function of baryon mass for any set of the
cosmological parameters $\Omega_m$, $\Lambda$, and $h$. The inner integral in
(\ref{eq:volume:m}) represents the effective value of the survey volume per
unit redshift interval for clusters of given mass, $(dV/dz)_{\text{eff}}$.

There is no simple scaling of $V(M)$ with either $\Omega_m$ or $\Lambda$.  In
addition to $dV/dz$, the index $A_{ML}$ is also cosmology-dependent.  When
fitting the mass function to different cosmological models, we recomputed
$V(M)$ for every new combination of the model parameters.

Figure~\ref{fig:mfun:volume} shows the comoving volume in the redshift
interval $0.4< z<0.8$ as a function of the baryon mass. Note that in the
limit of large masses, the survey volume approaches that of the local
all-sky surveys.

The right panel of Fig.~\ref{fig:mfun:volume} shows a volume per unit
redshift interval for three values of $M_b$. The clusters with small masses
generally have low luminosities and therefore can be detected only near the
lower boundary of the redshift interval considered. Massive clusters can be
detected at any $z$, and therefore most of the volume for such clusters is
accumulated near the higher boundary. This dependence of
$(dV/dz)_{\text{eff}}$ on mass should be taken into account in modeling
the mass function because the evolutionary effects between $z=0.4$ and $0.8$
are quite strong.

\begin{figure*}
 \centerline{\includegraphics[width=0.49\linewidth]{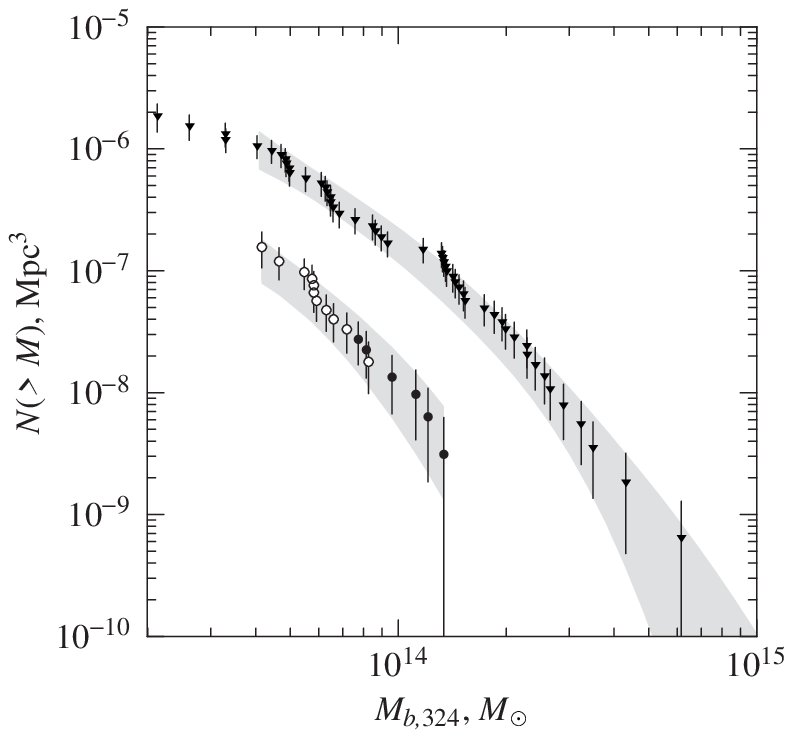}\hfill
   \includegraphics[width=0.49\linewidth]{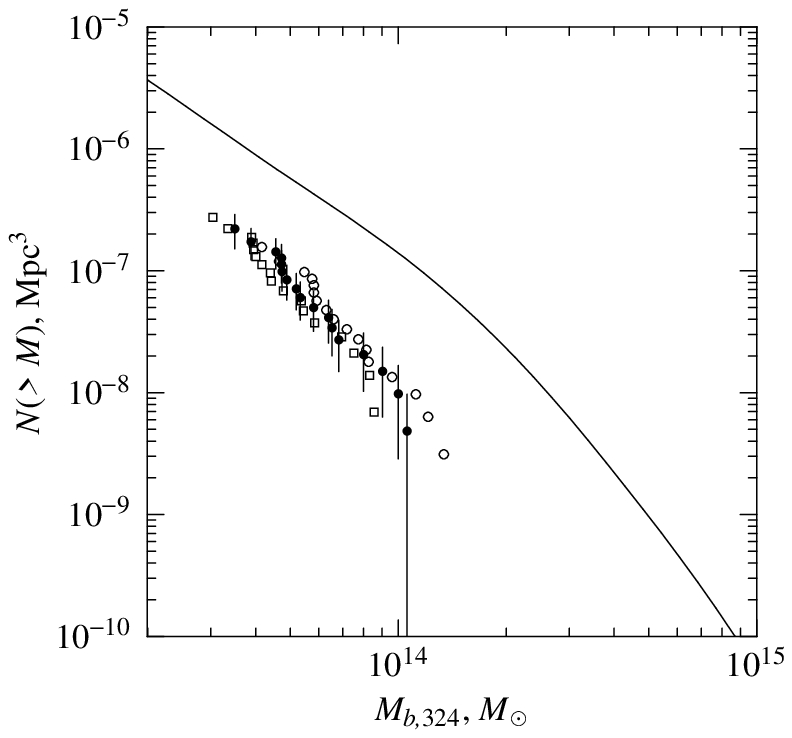}}
 \vskip -1.0\baselineskip
  \caption{\emph{Left:} The baryon mass function for the \160d\ survey sample in the
    redshift interval $0.4<z<0.8$. The local mass function from Voevodkin et
    al.\ (2002b) is shown by triangles. The grey band shows a 68\%
    uncertainty interval for the mass functions, including both the Poisson
    noise and mass measurement uncertainties. The mass function at $z>0.4$
    is shown by circles. The solid circles correspond to clusters with
    \emph{Chandra} baryon mass measurements, and the empty circles to the
    mass estimates from the $M_b-L$ relation.  The high-redshift mass
    function is computed for $\Omega_m=0.3$, $\Lambda=0.7$; the local mass
    function is cosmology-independent. \emph{Right:} Baryon mass functions
    derived assuming $\Omega_m=0.3$, $\Lambda=0.7$ \emph{(empty circles)};
    $\Omega_m=0.3$, $\Lambda=0$ \emph{(solid circles)}; $\Omega_m=1$,
    $\Lambda=0$ \emph{(squares)}. The solid line shows a fit to the local
    mass function. On both panels, the data are for $h=0.5$. The measured
    $M_b$ scales as $M_b\propto h^{-2.25}$ and volume scales as usual,
    $V\propto h^{-3}$}
  \label{fig:mfun:cosm}
  \label{fig:mfun}
\end{figure*}

\subsection{Observed Baryon Mass Function}

Since the number of clusters in our sample is rather small, we will work
with the integral representation of the mass function, which is determined
as
\begin{equation}
  F(M) = N(>M) = \sum_{M_b^{(i)}>M} \frac{1}{V(M_b^{(i)})},
\end{equation}
The Poisson errorbars for $F(M)$ are computed as
\begin{equation}
  (\Delta F)^2 = \sum_{M_{b}^{(i)}>M} \frac{1}{V^2(M_b^{(i)})}.
\end{equation}
In addition to the Poisson noise, the $F(M)$ errorbars should include an
additional component reflecting the individual mass measurement
uncertainties. This was done using the Maximum Likelihood analysis described
in the Appendix. We estimate that other obvious sources of uncertainty in
$F(M)$ are small. For example, the inaccuracies of the survey volume caused
by uncertainties in the \160d{} sky coverage or variations of the evolution
or scatter in the $M_b-L$ relation within the observationally allowed
intervals are all below 10\%, which is much smaller than the Poisson noise
in the mass function, so we ignore them.

The baryon mass function computed in the $\Omega_m=0.3$, $\Lambda=0.7$
cosmology is shown in Fig.~\ref{fig:mfun}. A comparison with the local
$F(M)$ from VVPb shows a strong evolution --- at $z\approx0.5$ there are
approximately 10 times fewer clusters of the same mass per unit comoving
volume than at $z=0$. This evolution is quite insensitive to the assumed
cosmology.  The right panel in Fig.~\ref{fig:mfun:cosm} shows the mass
functions computed assuming several combinations of $\Omega_m$ and
$\Lambda$. Mostly, the changes are restricted to a translation along the
band defined by $F(M)$ itself.

Part of the evolution of the mass function in Fig.~\ref{fig:mfun} may be
unrelated to the physical growth of clusters and instead related to the
commonly used definition of mass. Indeed, let us assume that all clusters
stay exactly the same at all redshifts. The operational definition of mass
used in our study is related to the density contrast above the mean density
of the Universe at the cluster redshift. Since the mean density changes as
$(1+z)^3$, the radius of the fixed contrast decreases with redshift, and
therefore the mass also decreases. For the average density profile at large
radii, we find that this effect results in the mass scaling $M\sim
(1+z)^{-1.5}$. This mass offset explains about 40\% of the observed effect,
and the rest is due to the real growth, by a factor of $\sim 2$ in mass, of
clusters between $z>0.4$ and $z=0$.

As was explained above, the total mass functions are related to the baryon
mass functions through a constant log shift along the mass axis, which does
not alter the vertical offset between the local and high-redshift
functions. The vertical offset is related to the growth of density
fluctuations between $z>0.4$ and $z=0$. Below, we use it to derive
constraints on the parameters $\Omega_m$ and $\Lambda$.

\section{Theory}
\label{sec:theory}

Our computations of the theoretical mass functions are based on the Jenkins
et al.\ (2001) universal fit to the mass functions in large-volume
cosmological simulations. Jenkins et al.\ find that the mass functions
expressed in terms of the \emph{rms} density fluctuations on mass scale $M$
are of a universal form:
\begin{equation}\label{eq:mfun:jenkins}
  f(\sigma)=-\frac{M}{\langle\rho\rangle}\,\frac{d\,n(M)}{d\ln\sigma^{-1}} = 
  A\, \exp\left(-|\ln\sigma^{-1}+b\,|^c\right),
\end{equation}
where $n(M)$ is the number of clusters with mass greater than $M$. The
universality means that the coefficients $A$, $b$, $c$ do not depend on the
cosmological parameters, nor on redshift. All the cosmological dependencies
are in the function $\sigma(M,z)$. The values of $A$, $b$, and $c$ depend,
however, on the definition of the cluster mass. For the definition in use
here --- mass corresponding to the spherical overdensity $\delta=324$
relative to the mean density of the Universe at the redshift of observation
--- Jenkins et al.\ provide the values
\begin{equation}\label{eq:mfun:jenkins:pars}
  A = 0.316, \quad b=0.67, \quad c=3.82.
\end{equation}

The function $\sigma(M,z)$ is the product of the present-day \emph{rms}
density fluctuations, $\sigma(M)$, and the growth factor of linear density
perturbations, $D(z)$. The growth factor as a function of $z$ is uniquely
determined by the values of the cosmological parameters $\Omega_m$ and
$\Lambda$. We used the software provided by Hamilton (2001) to compute
$D(z)$ in any cosmology. The function $\sigma(M)$ is trivially computed from
the present-day linear power spectrum, $P(k)$. We assume that $P(k)$ is the
product of the inflationary power law spectrum, $k^n$, and the transfer
function for the given mixture of CDM and baryons. The computation of the
transfer function for any set of cosmological parameters was performed using
the analytic approximations developed by Eisenstein \& Hu (1998) to the
exact numerical calculations with \textsc{cmbfast} (Seljak \& Zaldarriaga
1996).

Finally, the computation of the theoretical mass function according to
(\ref{eq:mfun:jenkins}) was performed using the software kindly provided by
A.~Jenkins. The program was slightly modified to include the power spectrum
model by Eisenstein \& Hu and the $D(z)$ computations due to Hamilton. With
these modifications, we have the machinery for precise computation of the
total cluster mass function for any combination of the cosmological
parameters at any redshift.

As was discussed above, if the baryon fraction in clusters equals its
universal value, the model baryon and total mass functions are related as
$F_b(M_b) = F(M_b\,\Omega_m/\Omega_b)$. Let us consider how this is modified
if the baryons in clusters are under-abundant by a factor $\Upsilon$: $M_b/M
= \Upsilon\, \Omega_b/\Omega_m$. In this case, the total mass that corresponds
to the measured baryon mass $M_{b,324}$, equals
$M_{324}=\Upsilon^{-1.5}\,\Omega_m/\Omega_b\,M_{b,324}$. The factor
$\Upsilon^{-1}$ here reflects the underabundance of baryons. The additional
factor $\Upsilon^{-0.5}$ follows from the fact that the total and baryonic
overdensities are now different. We would like to have the masses within the
radius of total overdensity $\delta=324$. This corresponds to the baryonic
overdensity $\delta_b = \Upsilon\,324$, but we measure $M_b$ within
$\delta_b = 324$. We find empirically that the density profile of a typical
cluster corresponds to the mass profile $M_{b,\delta}\propto \delta^{-0.5}$,
and hence the measured baryon masses have to be scaled approximately by the
factor $\Upsilon^{-0.5}$. Therefore, we have the following relation for the
theoretical model of the experimental baryon mass function:
\begin{equation}\label{eq:f(m_b):f(m):details}
  F_b(M_b) = F\left(\Upsilon^{-1.5}\,\Omega_m/\Omega_b\, M_b\right),
\end{equation}
where $F(M)$ is computed from equation (\ref{eq:mfun:jenkins}). 

\subsection{Dependence of the theoretical mass function on the
  model parameters}

Full details of our modeling procedure will be given below
(\S\,\ref{sec:fitting}). Here we present only a brief outline necessary to
understand the influence of the model parameters on the proposed
cosmological test.  Our measurements are the baryon mass functions at
$z\approx0$ and $z=0.4-0.8$. All the constraints will be obtained by the
requirement that the model mass functions are consistent with both the low-
and high-redshift measurements. In this Paper, we consider only constraints
from the cluster evolution and minimize the usage of any other
information. In particular, we will not fit the shape of the local baryon
mass function, but only use its normalization in the
$5\times10^{13}\,M_\sun<M_b<10^{14}\,M_\sun$ mass range
(approximately the median mass of the low-redshift sample). The above mass
interval corresponds to ICM temperatures of 3.7--5.5~keV.

The theoretical model for the baryon mass function depends on the following
parameters: $\Omega_m$; $\Lambda$; average baryon density, $\Omega_b\,h^2$;
the slope of the primordial power spectrum, $n$; present-day normalization
of the power spectrum, $\sigma_8$; Hubble constant, $h$; and also on any
deviations of the baryon fraction in clusters from the universal value. Let
us discuss how each of these parameters enters the model computations and
comparison of the model with observations.

\emph{Parameter} $n$:\quad The slope of the primordial power spectrum
determines the general slope of the mass functions. When $n$ varies, the
high-redshift and local mass functions change in unison. Since in the nearby
clusters, we consider only the overall normalization of the mass function,
and the measurement of the high-$z$ mass function has large statistical
uncertainties, our evolutionary test is insensitive to variations of $n$ in
the interval $0.5<n<1.5$ (as we explicitly verified). This is much wider
than the interval allowed by the current CMB observations, $n=1.05\pm0.1$
(Sievers et al.\ 2002, Rubi\~no-Martin et al.\ 2002). We use $n=1$ for our
baseline model.

\emph{Parameter} $h$:\quad The Hubble constant enters through its effect on
the shape of the power spectrum (mainly via the product $\Omega_m h$, Bond
\& Efstathiou 1984). It also affects the scaling between the total masses in
the model and the measured baryon masses of clusters. The effect of $h$ on
the power spectrum is unimportant for our test for the same reason, as that
for $n$. The effect of $h$ on the scaling between $M_b$ and $M$ is a shift
of the model mass function by a factor of $h^{0.75}$ with respect to the
measurements (for details, see Voevodkin \& Vikhlinin in preparation). In
the mass range of interest, the model mass functions at high $z$ and at
$z=0$ are almost parallel and therefore a shift along the mass axis can be
compensated by a slight change in $\sigma_8$. We have checked, that the
constraints on $\Omega_m$ and $\Lambda$ are unchanged for any $h$ in the
interval $0.5<h<0.8$. For the baseline model, we use $h=0.65$.

\emph{Parameter} $\Omega_b\,h^2$:\quad The average baryon density in the
Universe is given by BBN theory, $\Omega_b h^2 = 0.020\pm0.001$ (Burles et
al.\ 2001). A larger variation of this parameter would cause a slight change
in the shape of the present-day power spectrum, and also would affect the
scaling between the total and baryon mass. As was already noted, these
changes are unimportant for our evolutionary test. We have verified that our
results are unchanged for any value in the interval
$0.015<\Omega_b\,h^2<0.025$. In the baseline model, we use $\Omega_b
h^2=0.02$.

\emph{Parameter} $\sigma_8$:\quad The normalization of the mass functions is
exponentially sensitive $\sigma_8$. The requirement that the theoretical
baryon mass function fits the measurements at $z=0$ effectively fixes
$\sigma_8$ for any combination of other model parameters. The measurement of
$\sigma_8$ by this technique will be presented elsewhere. For the purposes
of this study, all our constraints are marginalized over the acceptable
(i.e., those consistent with the local mass function) values of $\sigma_8$.

\emph{Parameters} $\Omega_m$ and $\Lambda$:\quad
The cosmological density parameter determines (mainly via the product
$\Omega_m h$) the slope of the power spectrum on cluster scales, and hence the
shape of the mass function. By design, our test is rather insensitive to the
shape of the mass function (see discussion of the parameter $n$ above).

In addition, a combination of parameters $\Omega_m$ and $\Lambda$ determines
the growth factor of linear density perturbations, and therefore the
evolution of the mass functions between high redshift and $z=0$. This effect
is the basis for our evolutionary test.

\emph{Deviations of the baryon fraction from universality.}\quad Any
deviations of the cluster baryon fraction from the universal value,
$\Omega_b/\Omega_m$, can potentially be a serious problem for cosmological
tests using the baryon mass function. There are three possibilities: 1)
$\Upsilon \ne 1$, but is the same for all clusters; 2) $\Upsilon=1$ on
average, but there is a cluster-to-cluster scatter; 3) $\Upsilon$ is a
function of cluster mass (e.\,g.\ $\Upsilon$ becomes small for low-mass
clusters). Taking these possibilities into account is important for tests
based on the shape of the baryon mass function and for the measurement of
$\sigma_8$ (a detailed discussion will be given in Voevodkin \&
Vikhlinin). However, the evolutionary test discussed here is rather
insensitive to $\Upsilon$, as long as there is no strong evolution of this
parameter with redshift, as shown below.

Let $\Upsilon\ne 1$, but be the same for all clusters. This could be due to
different virialization processes for the dark matter and the baryon
gas. Adiabatic numerical simulations of cluster formation often lead to
$\Upsilon = 0.9-0.95$, independent of the cluster mass (Mathiesen, Mohr \&
Evrard 1999). That paper also considers the systematic bias in the X-ray
measurements of the baryon mass due to deviations of cluster shapes from
spherical symmetry, which for our purposes is indistiguishable from
$\Upsilon\ne1$. The Mathiesen et al.\ results suggest that deviations from
spherical symmetry are equivalent to $\Upsilon=1.02-1.12$, almost
independent of mass. The effect of a constant $\Upsilon\ne1$ on our
evolutionary test is to shift all mass functions along the $M$ axis by
$\Upsilon^{-1.5}$ (cf.\ eq.~\ref{eq:f(m_b):f(m):details}). As was already
discussed, this shift is unimportant. We verified that the obtained
cosmological constraints are virtually identical for $\Upsilon=0.85$, 1, and
1.15.

If there is scatter in $\Upsilon$, the model for the baryon mass function
should be convolved with some kernel. As long as the scatter is not too
large (the observational upper limits on the variations of $f_b$ are
approximately 15--20\%, see, e.g., Mohr et al.\ 1999), its effect on the
mass function is very small.
%  Morever, the scatter $\Upsilon$ for the nearby and
% distant clusters can be expected the same at least in the first
% approximation. In this case the effect on the distant and low-redshift mass
% function would be identical and therefore would not affect the evolutionary
% test. 

Let us now consider the possibility of a systematic change of $\Upsilon$
with the cluster mass. For example, a strong preheating of the intergalactic
medium can cause the baryons to be underabundant in low-mass clusters
(Cavaliere, Menci \& Tozzi 1997). Bialek et al.\ (2001) discuss the
cosmological simulations of clusters with the level of preheating adjusted
to reproduce the local $L_x-T$ relation. They find a systematic decrease of
$\Upsilon$ in low-temperature clusters, which can be written in terms of the
baryon mass approximately as:
\begin{equation}\label{eq:Upsilon:M:bialek}
  \Upsilon = 1-M_*/M_b, \quad M_* = 0.9\times10^{13}\,M_\sun.
\end{equation}
Such a dependence of $\Upsilon$ on mass distorts the shape of the
experimental baryon mass function quite significantly, but mainly for the
low mass clusters which we do not use here. We used
eq.~(\ref{eq:Upsilon:M:bialek}) in our baseline model, but have verified
that with $\Upsilon=1$, the results are virtually identical.

Finally, we also considered a possible evolution of
eq.~(\ref{eq:Upsilon:M:bialek}) with redshift. A reasonable assumption is
that $M_*$ in (\ref{eq:Upsilon:M:bialek}) corresponds to the fixed ratio of
the entropy of the intracluster and intergalactic media, which leads to the
scaling $M_*\propto (1+z)^{3/2}$. We verified that the constraints on
$\Omega_m$ and $\Lambda$ are very weakly affected by such a scaling.

The evolution of $\Upsilon$ is important for our test only if it is
significant for massive clusters. The amplitude of the statistical
uncertainties on $\Omega_m$ and $\Lambda$ is equivalent to a $\pm13\%$ change
in $\Upsilon$ at $z=0.55$. We consider such changes unlikely because the
physically motivated value for massive clusters is $\Upsilon=1$.

\section{Fitting procedure}
\label{sec:fitting}

The tightest parameter constraints can be obtained by the maximum-likelihood
analysis of the distribution of clusters as a function of both mass and
redshift. However, given the novelty of our cluster evolution test, we
decided to model the two experimental mass functions, at $z\approx0$ and
$0.4<z<0.8$, using goodness of fit information and rejecting only those
models which are clearly inconsistent with the data. This is a simpler and
more convincing method then the maximum-likelihood fit. Our approach is
described below.

While the computation of the model at $z=0$ is straightforward
(eq.~\ref{eq:mfun:jenkins},~\ref{eq:mfun:jenkins:pars},~\ref{eq:f(m_b):f(m):details},~\ref{eq:Upsilon:M:bialek}
at $z=0$), it is less so for our high-$z$ sample because one expects a
significant evolution within the redshift interval considered,
$0.4<z<0.8$. To take this into account, we subdivided this interval into 10
narrow $z$ bins, computed cumulative mass functions on the same grid of
masses in each bin, and then weighted them with $(dV/dz)_{\text{eff}}$ (see
eq.~\ref{eq:volume:m} and Fig.~\ref{fig:mfun:volume}) for each mass. This
gives the model mass function which can be directly compared with the one
observed at $0.4<z<0.8$.

Note that the volume computations depend on the evolution of the $M_b-L$
relation (cf.\ eq.~\ref{eq:volume:m}), which is slightly
cosmology-dependent. For self-consistency, for each combination of $\Omega_m$
and $\Lambda$ we recomputed masses and luminosities of all clusters in the
Vikhlinin et al.\ (2002) sample and rederived the value of the index
$A_{ML}$ (cf.\ eq.~\ref{eq:volume:m:ml}). This index also enters the volume
computations for the experimental baryon mass function, which we also
recomputed for each set of parameters $\Omega_m$, $\Lambda$, and $h$.

As was discussed above, our test is insensitive to the precise values of the
cosmological parameters $n$, $\Omega_b$, and $h$. Therefore, we fixed them
at $n=1$, $h=0.65$, and $\Omega_b = 0.02\,h^2$, and varied only $\Omega_m$,
$\Lambda$, and $\sigma_8$. The model mass functions were computed on a grid
$\Omega_m=0.05,0.075,\ldots,1.25$, $\Lambda=0.0,0.05,\ldots,1.1$, and
$\sigma_8=0.4,0.405,\ldots,1.5$. The constraint on $\Omega_m$ and $\Lambda$
was marginalized over $\sigma_8$ --- a combination of $\Omega_m$ and $\Lambda$
was considered acceptable if for some value of $\sigma_8$, the model mass
functions were consistent with the local and high-redshift measurements
simultaneously.

The comparison of the theoretical mass functions with observations was
designed to rely mainly on the observed evolution and to limit the use of
any other information. In particular, we use only the normalization of the
local mass function in the mass range
$5\times10^{13}\,M_\sun<M_b<1\times10^{14}\,M_\sun$ (for $h=0.65$).  In the distant
sample, we will consider the $F(M)$ uncertainty intervals in the identical
mass range.

A simple and conservative comparison of the theoretical and observed mass
functions is performed as follows. If the model cumulative function is
entirely within the 68\% confidence interval of the observed one, it is
considered acceptable at the 68\% confidence level. The confidence levels of
90\% and 95\% are applied identically.  The combined confidence level for
the local and distant mass functions are combined according to
Table~\ref{tab:mfun:fitting:sign}. For example, a model is acceptable at the
90\% confidence level if it is within the 68\% error bars of one of the mass
functions, and within the 90\% error bars of another. It can be shown that
the described method leads to more conservative (wider) confidence intervals
compared to the likelihood ratio test.

\begin{table}
  \caption{Combined confidence level for the local and high-redshift mass functions}\label{tab:mfun:fitting:sign}
\footnotesize\def\arraystretch{1.5} \setlength{\unitlength}{1.5mm}
\centerline{
\begin{tabular}{|c|c|c|c|}
\hline
\begin{picture}(10.5,5)(0,0)
\put(0,0){$z>0.4$}
\put(-0.4,5){\line(2,-1){12.0}}
\put(7,3){$z=0$}
\end{picture}
 & \raisebox{7pt}{68\%} & \raisebox{7pt}{90\%} & \raisebox{7pt}{95\%} \\
\hline
68\%         & 68\% & 90\% & 95\% \\
\hline
90\%         & 90\% & 95\% & --   \\
\hline
95\%         & 95\% & --   & --   \\
\hline
\end{tabular}
}
\end{table}

\section{Results}
\label{sec:omega:lambda:results}

\begin{figure*}
  \centerline{
    \includegraphics[width=0.49\linewidth]{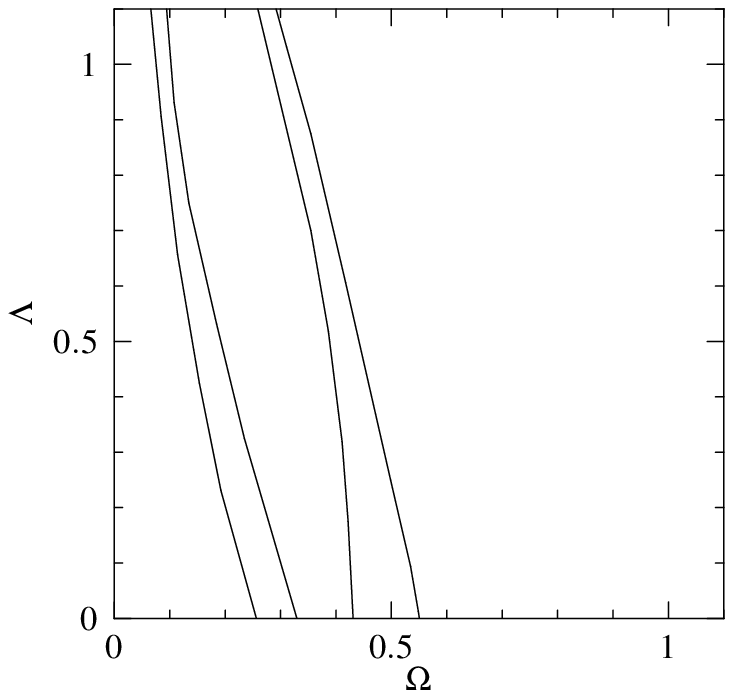}
    \includegraphics[width=0.49\linewidth]{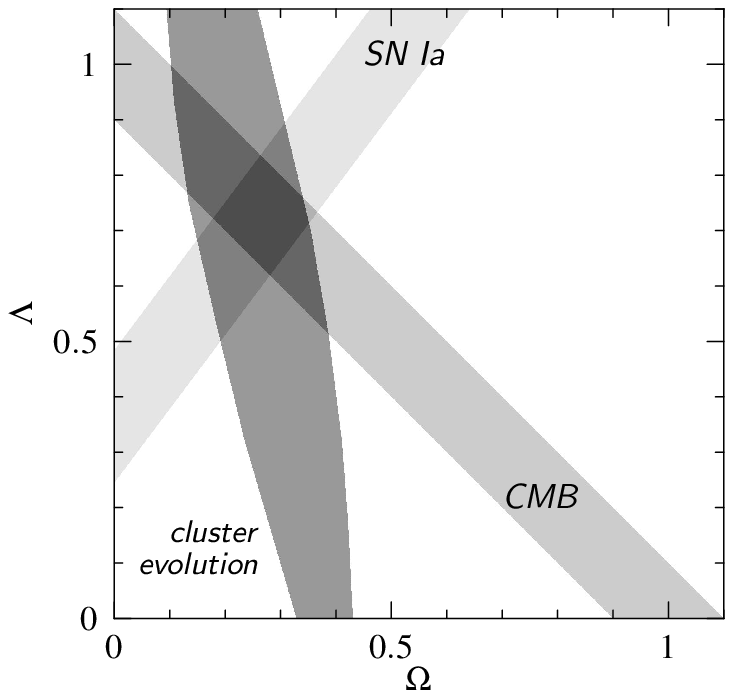}
  }
  \vskip -1.0\baselineskip
  \caption{\emph{Left:} 68\% and 90\% confidence intervals on $\Omega_m$ and 
    $\Lambda$ from evolution of the baryon mass function at $z>0.4$.
    \emph{Right:} Comparison of the cluster evolution constraints with those
    from the distant supernovae and CMB. For each technique, we show only
    68\% confidence regions.  }
  \label{fig:cosm:evol}
  \label{fig:cosm:evol:cmb:sn}
\end{figure*}

The resulting 68\% and 90\% confidence regions for parameters $\Omega_m$ and
$\Lambda$ are shown in Fig.~\ref{fig:cosm:evol}. The observed rate of the
evolution of the baryon mass function defines a narrow band in the
$(\Omega_m,\Lambda)$ plane, slightly inclined to the $\Lambda$ axis:
\begin{equation}
  \Omega_m+0.23\,\Lambda= 0.41 \pm 0.10.
\end{equation}
The uncertainty in $\Omega_m$ for any fixed $\Lambda$ is $\pm0.10$ at the 68\%
confidence level. For a flat cosmology, $\Omega_m+\Lambda=1$, we have
$\Omega_m=0.24\pm0.12$. Using \emph{only} clusters with direct \emph{Chandra}
measurements of $M_b$, we obtain identical constraints on $\Omega_m$ and
$\Lambda$, but the uncertainties are larger by a factor of $\sim 1.3$.

The band of acceptable values of $\Omega_m$ and $\Lambda$ approximately
corresponds to a constant growth factor of linear density perturbations
between $z\approx0.55$ and $z\approx0.05$, corresponding to the observed
factor of 10 evolution of the mass function. For small values of $\Omega_m$, a
significant growth factor is achieved only if $\Lambda$ is large (Carrol et
al.\ 1992). For $\Lambda=0$ the required growth factor is achieved for
$\Omega_m\approx 0.4$.

As has become a common practice, we compare our $(\Omega_m,\Lambda)$
constraints with those provided by the observations of distant type Ia
supernovae and the CMB fluctuations. The most robust constraint from the
current CMB observations follows from the location of the first Doppler
peak, which indicates that the Universe is nearly flat,
$\Omega_{\text{tot}}=\Omega_m+\Lambda=1\pm0.1$ (with a $0.4<h<0.9$ prior;
Sievers et al.\ 2002, Rubi\~no-Martin et al.\ 2002). The distant supernovae
data require a positive cosmological constant. We combine the SN~Ia
constraints from Riess et al.\ (1998) and Perlmutter et al.\ (1999):
$\Lambda-1.33\,\Omega_m=0.37\pm0.12$.  The corresponding 68\% confidence
intervals are shown in the right panel of
Fig.~\ref{fig:cosm:evol:cmb:sn}. Remarkably, the constraints from all three
methods intersect in a small region, $\Omega_m=0.27\pm0.085$,
$\Lambda=0.73\pm0.11$.

\section{Discussion and conclusions}

We have presented a new method of applying the models of the cluster
evolution to observations at high redshifts. This method is based on easily
observable baryon mass function. Unlike the studies based on the cluster
temperature or luminosity functions, our method does not suffer from large
uncertainties in the $M_{\text{tot}}-T$ and $M_{\text{tot}}-L$ relations. It
uses only a physically motivated assumption that the baryon fraction in the
cluster mass within the large radii is close to the universal value,
$\Omega_b/\Omega_m$.  Systematic uncertainties in this method are smaller
and very different from those using the temperature or luminosity
functions. Until improved observational techniques are able to normalize the
scaling relations to a $10\%$ accuracy at all redshifts, modeling of the
baryon mass function appears to be a promising method for deriving the
cluster-based cosmological constraints.

The application of this method to the subsample of distant clusters from the
\160d\ survey results in a tight constraint on the combination of
cosmological parameters $\Omega_m$ and $\Lambda$. This constraint intersects
the current CMB and SN~Ia constraints in a small area near $\Omega_m=0.3$ and
$\Lambda=0.7$, showing a remarkable agreement of totally independent
methods.

We note an excellent agreement of our cluster-derived cosmological parameter
$\Omega_m$ with the results based on the X-ray measurements of the total mass
and baryon fraction in the intermediate-redshift clusters (Allen et al.\
2002). Also, there is a good agreement with those studies of the XTF
evolution in the EMSS sample which use the normalization of the $M-T$
relation based on the X-ray mass measurements (e.g., Henry 2000). Such
consistency indirectly confirms the validity of the X-ray cluster mass
measurements.

Our method relies on the assumption that the baryon fraction in massive
clusters does not evolve with redshift. This assumption is well-supported by
numerical simulations. One issue of concern, however, is that these same
simulations do not reproduce the observed evolution of the $M_b-T$
relation. At a fixed $T$, we observe $M_b \propto (1+z)^{-0.5\pm0.4}$ for
$\Omega_m=0.3$ and $\Lambda=0.7$ (Vikhlinin et al.\ 2002), which is
significantly different from the expected evolution $(M_b,
M_{\text{tot}})\propto T^{3/2}(1+z)^{-3/2}$ (e.g., Bryan \& Norman 1998).
This may indicate that the baryon fraction in clusters evolves with
redshift. However, it is much more likely that simulations simply do not
reproduce the distribution of gas in the cluster central regions. These
regions usually dominate in the observed emission-weighted $T$ but at the
same time the gas there is especially prone to processes such as radiative
cooling which can easily change $T$ but are unlikely to modify $M_b$ or
$M_{\text{tot}}$.

The accuracy of our cosmological constraints is presently limited by the low
number of the distant clusters. When new distant cluster samples from the
\emph{Chandra}, \emph{XMM}, and extended \emph{ROSAT} surveys become
available, and more distant clusters are observed with adequate exposures by
\emph{Chandra} and \emph{XMM}, even tighter cosmological constraints will be
possible. Eventually, different methods (cluster evolution, CMB, SN~Ia etc.)
will start to disagree, which will lead to better understanding of the
systematics and warrant more complicated theoretical models, e.g.\ those
involving non-standard equations of state for the dark energy.

\acknowledgements

We thank A.~Jenkins for providing the software for computing the model mass
functions. Useful discussions with M.~Markevitch and A.~Kravtsov are
gratefully acknowledged. A.~Voevodkin thanks SAO for hospitality during the
course of this research. This work was supported by NASA grant NAG5-9217 and
contract NAS8-39073. HQ is partially supported by FONDAP Centro de
Astrofisica (Chile).

\begin{appendix}
  \section{Maximum likelihood fitting of the observed mass functions}
  The uncertainty intervals on the low- and high-redshift mass functions
  were obtained by the Maximum Likelihood analysis of the unbinned mass
  measurements. Our technique is presented in VVPb and briefly summarized
  below.

  We assume that the observed mass function can be adequately described by
  the Schechter model,
  \begin{equation}\label{eq:schechter}
    f(M) = dN/dM = A M^{-\alpha} \exp (-M/M_*),
  \end{equation}
  in the narrow mass ranges of interest. The variation of parameters $A$,
  $\alpha$, and $M_*$ within the 3-parameter $N\%$ confidence intervals
  produces a band of the cumulative mass function models which we associate
  with the $N\%$ confidence band for the observed cumulative mass
  function. 

  The confidence intervals for parameters $A$, $\alpha$, and $M_*$ are
  obtained from the Maximum Likelihood modeling of the unbinned observed
  mass measurements. In the case of negligible mass measurement errors, the
  likelihood function is (cf.\ Cash 1979):
  \begin{equation}\label{eq:likelihood}
    \log L = \sum_i \log\left(f(M_i) V(M_i)\right) - \int f(M) V(M)\,dM,
  \end{equation}
  where $M_i$ are the individual mass measurements, $V(M)$ is the sample
  volume as a function of mass (computed as described in
  \S\,\ref{sec:mfun:volume}), and the integral is over the entire mass
  range. The quantity $C=-2\log L$ is statistically equivalent to $\chi^2$
  (Cash 1979), therefore the uncertainties of the estimated parameters can
  be derived from the standard $\Delta \chi^2$ test. For example, a combined
  68\% confidence region for parameters $A$, $\alpha$, and $M_*$ is defined
  by the deviation $\Delta C = 3.5$ from the minimum (the 68\% probability
  point for the $\chi^2$ distribution with 3 degrees of freedom).

  In the case of the finite mass measurement errors, the product $f(M) V(M)$
  in~(\ref{eq:likelihood}) should be convolved with the distribution of the
  measurement scatter. We assume that the scatter follows the log-normal
  distribution,
  \begin{equation}
   \frac{d P(\log M)}{d\log M} = \frac{\exp (-(\log M - \log
      \mu)^2/2\sigma^2)}{(2\pi)^{1/2}\sigma},
  \end{equation}
  where $\mu$ is the true mass. The convolution of $f(M) V(M)$ with this
  distribution is
  \begin{equation}
    F(M,\sigma) = \int \frac{\exp (-(\log M - \log
      \mu)^2/2\sigma^2)}{(2\pi)^{1/2}\sigma} f(\mu) V(\mu) \,d\log\mu
  \end{equation}
  and the likelihood function is
  \begin{equation}
    \log L = \sum_i \log\left(F(M_i,\sigma_i)\right) - \int_M F(M,\sigma(M))
    \,dM,
  \end{equation}
  where $\sigma_i$ is the log scatter of the $i$-th measurement and
  $\sigma(M)$ is the trend of the typical measurement scatter with mass.

  The 68\% confidence intervals for the cumulative mass function obtained
  with this technique are shown by grey bands in
  Fig.~\ref{fig:mfun:cosm}. For the low-redshift mass function, the mass
  measurement errors are small and we recover the pure Poisson scatter. The
  derived confidence interval for the distant mass function is wider than
  the Poisson scatter because the masses are less accurate.

  The described technique automatically corrects any biases in the mass
  function determination which could arise from the measurement
  uncertainties (e.g., the Eddington bias). The close agreement between the
  model (not convolved with the measurement errors) and the data in
  Fig.~\ref{fig:mfun:cosm} shows that such biases are small in our case. A
  bias could be expected because the mass function is steep, but is absent
  because of the rather flat distribution of the observed number of clusters
  in our sample per log interval of mass (this is the product of the
  intrinsic mass function and the selection function). The effect of the
  mass measurement uncertainties on the mass function is to smooth the
  distribution of $d N/d\log M$. If this distribution is flat, the net
  effect of the measurement uncertainties is zero. We verified the absence
  of biases in our case using Monte-Carlo simulations.

\end{appendix}


\begin{references}

  \reference{2002MNRAS.334L..11A} Allen, 
  S.~W., Schmidt, R.~W., \& Fabian, A.~C.\ 2002, \mnras, 334, L11 

  \reference{2001ApJ...555..597B} Bialek, J.~J., Evrard, A.~E., \& Mohr,
  J.~J.\ 2001, \apj, 555, 597 

  \reference{2000A&A...362..809B} Blanchard, A., Sadat, R., Bartlett, J.~G.,
  \& Le Dour, M.\ 2000, \aap, 362, 809

  \reference{1984ApJ...285L..45B} Bond, J.~R.~\& Efstathiou, G.\ 1984, \apjl,
  285, L45 

  \reference{2001ApJ...561...13B} Borgani, S.~et al.\ 2001, \apj, 561, 13 

  \reference{1998ApJ...495...80B} Bryan, G.~L. \& Norman, M.~L. 1998, ApJ,
  495, 80

  \reference{2001ApJ...552L...1B} Burles, S., 
  Nollett, K.~M., \& Turner, M.~S.\ 2001, \apjl, 552, L1 

  \reference{1992ARA&A..30..499C} Carroll, S.~M., Press, W.~H., \& Turner,
  E.~L. 1992, ARA\&A, 30, 499 

  \reference{1979ApJ...228..939C}  Cash, W. 1979, ApJ, 228, 939 

  \reference{1997ApJ...484L..21C} Cavaliere, A., Menci, N., \& Tozzi, P.\
  1997, \apjl, 484, L21 

  \reference{1999ApJ...523L.137D} Donahue, M.~\& Voit, G.~M.\ 1999, \apjl,
  523, L137

  \reference{2000ApJ...534..133E} Ebeling, H.~et al.\ 2000, \apj, 534, 133 

  \reference{1998ApJ...496..605E} Eisenstein, D.~J.~\& Hu, W.\ 1998, \apj,
  496, 605 

  \reference{1998MNRAS.298.1145E} Eke, V.~R., Cole, S., Frenk, C.~S., \&
  Henry, J. P.\ 1998, \mnras, 298, 1145

  \reference{1989ApJ...341L..71E} Evrard, A.~E.\ 1989, \apjl, 341, L71

  \reference{2002ApJ...573....7E} Evrard, A.~E.~et al.\ 2002, \apj, 573, 7


  \reference{2001A&A...368..749F} Finoguenov, A., Reiprich, T.~H., \&
  B{\"o}hringer, H.\ 2001, \aap, 368, 749


  \reference{1997AJ....114...14F} Fischer, P.~\& Tyson, J.~A.\ 1997, \aj, 114, 14 

  \reference{1998ApJ...503..518F} Fukugita, M., Hogan, C. J. \& Peebles,
  P. J. E. 1998, ApJ, 503, 518

  \reference{} Gioia, I.\ M., Maccacaro, T., Schild, R.\ E., Wolter, A.,
  Stocke, J.\ T., Morris, S.\ L., \& Henry, J.\ P.\ 1990,
  ApJS, 72, 567

  \reference{2001ApJ...553L.105G} Gioia, I.~M., Henry, 
  J.~P., Mullis, C.~R., Voges, W., Briel, U.~G., B{\" o}hringer, H., \& 
  Huchra, J.~P.\ 2001, \apjl, 553, L105 

  \reference{2001MNRAS.322..419H} Hamilton, A.~J.~S.\ 2001, \mnras, 322, 419


  \reference{1991ApJ...372..410H} Henry, J.~P.~\& Arnaud, K.~A.\ 1991, \apj,
  372, 410

  \reference{} Henry, J.\ P., Gioia, I.\ M., Maccacaro, T., Morris, S.\ L.,
  Stocke, J.\ T., \& Wolter, A.\ 1992, ApJ, 386, 408

  \reference{1997ApJ...489L...1H} Henry, J.~P.\ 1997, \apjl, 489, L1 

  \reference{2000ApJ...534..565H} Henry, J.~P.\ 2000, \apj, 534, 565 


  \reference{1999ApJ...520...78H} Horner, D.~J., Mushotzky, R.~F., \&
  Scharf, C.~A.\ 1999, \apj, 520, 78

  \reference{2001MNRAS.321..372J} Jenkins, A., Frenk, 
  C.~S., White, S.~D.~M., Colberg, J.~M., Cole, S., Evrard, A.~E., Couchman, 
  H.~M.~P., \& Yoshida, N.\ 2001, \mnras, 321, 372 


  \reference{2002ApJ...566..744L} Lewis, A.~D., Stocke, J.~T., Ellingson, E.,
  \& Gaidos, E.~J.\ 2002, \apj, 566, 744 



  \reference{1997ApJ...491..467M}  Markevitch, M. \& Vikhlinin, A. 1997,
  ApJ, 491, 467 

  \reference{1999ApJ...520L..21M} Mathiesen, B., Evrard, A.~E., \& Mohr,
  J.~J.\ 1999, \apjl, 520, L21 


  \reference{2000ApJ...532..694N} Nevalainen, J., Markevitch, M., \& Forman,
  W.\ 2000, \apj, 532, 694 

  \reference{1997ApJ...481..644N} Nichol, R.~C., Holden, 
  B.~P., Romer, A.~K., Ulmer, M.~P., Burke, D.~J., \& Collins, C.~A.\ 1997, 
  \apj, 481, 644 


  \reference{1980lssu.book.....P} Peebles, P. J. E. 1980, The Large-Scale
  Structure of the Universe (Princeton Univ. Press)

  \reference{1997NewA....2..309P} Pen, U.\ 1997, New Astronomy, 2, 309 

  \reference{2002ApJS..140..265P} Perlman, E.~S., Horner, D.~J., Jones, L.~R.,
  Scharf, C.~A., Ebeling, H., Wegner, G., \& Malkan, M.\ 2002, \apjs, 140, 265

  \reference{1999ApJ...517..565P} Perlmutter, S.~et al.\ 1999, \apj, 517, 565 


  \reference{}  Press, W.~H. \& Schechter, P. 1974, ApJ, 187, 425

  \reference{1999ApJ...518..521R} Reichart, D.~E., Nichol, R.~C., Castander,
  F.~J., Burke, D.~J., Romer, A.~K., Holden, B.~P., Collins, C.~A., \& Ulmer,
  M.~P.\ 1999, \apj, 518, 521

  \reference{2002ApJ...567..716R} Reiprich, T.~H.~\& B{\" o}hringer, H.\
  2002, \apj, 567, 716 

  \reference{1998AJ....116.1009R} Riess, A.~G.~et al.\ 1998, \aj, 116, 1009 

  \reference{} Rosati, P., Della Ceca, R., Norman, C., \& Giacconi, R.\ 1998,
                ApJ, 492, L21

  \reference{} Rosati, P., Borgani, S. \& Norman, C.\ 2002, ARA\&A, 40, 539

  \reference{} Rubi\~no-Martin, J.\ A.\ et al.\ 2002, MNRAS (submitted,
  astro-ph/0205367)

  \reference{1996PASJ...48L.119S} Sasaki, S.\ 1996, \pasj, 48, L119 

  \reference{1996ApJ...469..437S} Seljak, U.~\& Zaldarriaga, M.\ 1996, \apj,
  469, 437 

  \reference{} Sievers, J. L.\ 2002, ApJ (submitted; astro-ph/0205387)

  \reference{1971A&A....12..190S} Sunyaev, R.~A.\ 1971, \aap, 12, 190 

  \reference{1998ApJ...502..558V} Vikhlinin, A., McNamara, B.~R., Forman, W.,
  Jones, C., Quintana, H., \& Hornstrup, A.\  1998, \apj, 502, 558 

  \reference{1999ApJ...525...47V} Vikhlinin, A., Forman, W., \& Jones, C.\
  1999, \apj, 525, 47 


  \reference{2000lssx.proc...31V} Vikhlinin, A.~et al.\ 2000, in Proceedings
  of Large Scale Structure in the X-ray Universe, eds.~Plionis, M.~\&
  Georgantopoulos, I., (Atlantisciences, Paris)

  \reference{2002ApJ...578L.107V} Vikhlinin, A., 
  VanSpeybroeck, L., Markevitch, M., Forman, W.~R., \& Grego, L.\ 2002, 
  \apjl, 578, L107 

  \reference{2002AstL...28..366V} Voevodkin, A.~A., Vikhlinin, A.~A., \&
  Pavlinsky, M.~N.\ 2002a, Astronomy Letters, 28, 366 (VVPa)

  \reference{} Voevodkin, A.~A., Vikhlinin, A.~A., \&
  Pavlinsky, M.~N.\ 2002b, Astronomy Letters, 28, 793 (VVPb)

  \reference{1993Natur.366..429W} White, S. D.~M., Navarro, J.~F., Evrard,
  A.~E., \& Frenk, C.~S. 1993, Nature, 366, 429

\end{references}
\end{document}